\documentclass[twocolumn,floats,aps]{revtex4-1}
\usepackage{latexsym, amssymb, amsmath, titlesec, mathtools, hyperref, braket, accents}

\DeclareMathOperator{\tr}{tr}
\DeclareMathOperator{\Tr}{Tr}

\newcommand{\dif}{\text{d}}
\newcommand{\fdif}{\text{D}}

\newcommand{\Z}{\mathbb{Z}}

\newcommand{\fL}{\mathcal{L}}

\newcommand{\fP}{\mathcal{P}}
\newcommand{\cl}{\mathrm{c}}
\newcommand{\q}{\mathrm{q}}

\newlength{\dhatheight}
\newcommand{\doublehat}[1]{
    \settoheight{\dhatheight}{\ensuremath{\hat{#1}}}
    \addtolength{\dhatheight}{-0.35ex}
    \hat{\vphantom{\rule{1pt}{\dhatheight}}
    \smash{\hat{#1}}}}
\newcommand{\cev}[1]{\reflectbox{\ensuremath{\vec{\reflectbox{\ensuremath{#1}}}}}}

\begin{document}
\title{Qubit Decoherence and Symmetry Restoration through Real-Time Instantons}

\author{Foster Thompson$^1$
and Alex Kamenev$^{1,2}$}
\affiliation{$^1$School of Physics and Astronomy, University of Minnesota, Minneapolis, Minnesota 55455, USA}
\affiliation{$^2$William I. Fine Theoretical Physics Institute, University of Minnesota, Minneapolis, Minnesota 55455, USA}

\begin{abstract}
A parametrically driven quantum oscillator, stabilized by a nonlinear dissipation, exhibits a spontaneous breaking of the parity symmetry. It results in the quantum bi-stability, corresponding to a Bloch sphere of dark states. This makes such a driven-dissipative system  an attractive candidate for a qubit. The parity symmetry breaking is exact both on the classical level and within the quantum mechanical perturbation theory.
Here we show that non-perturbative quantum effects lead to the symmetry restoration and result in exponentially small but finite qubit decoherence rate.
Technically the symmetry restoration is due to real time instanton trajectories of the Keldysh path integral, which represents the Lindbladian 
evolution of the driven-dissipative oscillator.
\end{abstract}
\maketitle

Open quantum systems subject simultaneously to dissipative effects and a strong external drive obey effective dynamics described by the Lindblad equation~\cite{OpenQM,QuantumNoise,DrivenDissipativeLight,DrivenDissipativeColdAtoms,DiehlQuantumSim,Diehl}.  Such systems have gained attention for their ability to realize multiple non-equilibrium steady states~\cite{NonAbelianSym,AlbertGeom,NonEqSSKeldysh,AlbertSym,LindbladSymBreak,Prosen,Clerk,MultiDarkSpace}.  Of particular interest are systems exhibiting quantum bi-stability, in which the space of steady states has the structure of a Bloch sphere and can be used to encode a qubit stably in contact with its environment~\cite{QCompDrivenDissipative,AlbertCats,Albert}.

The stationary states of a Lindblad equation are determined by its symmetries \cite{AlbertSym,NonAbelianSym}; the spontaneous breaking of symmetry can lead to bi-stability \cite{LindbladSymBreak}.  A broken `strong' symmetry (in the sense of Ref.~\cite{Prosen}) leads to a  Bloch sphere stationary space \cite{Albert}, forming a qubit.  Similarly to the unitary quantum mechanics, non-perturbative effects often restore the symmetry, leading to a slow precession and decoherence of the would-be qubits. The quantum mechanical symmetry restoration (i.e. tunneling) is achieved through the instanton solution of the imaginary time equations of motion (EoM). Remarkably, the strong symmetry restoration in the dissipative Lindbladian dynamics occurs through {\em real time} solutions of complexified EoM. Consequently, this purely quantum phenomenon is akin to classical activation in a multidemnsional space~\cite{DykmanClassical1,DykmanClassical2}.

To exemplify this, consider a harmonic oscillator subject to an external drive that periodically modulates its resonant frequency $\hat H=\omega_0\hat a^\dagger\hat a+2\lambda\cos(2\omega_pt)(\hat a^{\dagger2}+\hat a^2)$, where $\hat a$ is a Bose annihilation operator, $[\hat a,\hat a^\dagger]=1$.  In addition to the coherent drive we allow the oscillator to dissipate through interaction with its environment, modelled as a bath of linear oscillators in thermal equilibrium at zero temperature.  To ensure the $\Z_2$ parity symmetry $\hat a\to\hat P\hat a\hat P^\dagger=-\hat a$ is still present when the oscillator is put in contact with the environment, we choose a nonlinear coupling between the system and the bath: $\hat H_\mathrm{int}=\sum_s\frac{g_s}{\sqrt{2\omega_s}}(\hat b^\dagger_s\hat a^2+\hat a^{\dagger2}\hat b_s)$, where bath modes $\hat b_s$ oscillate at frequencies $\omega_s$.

In the rotating frame of the external drive, ignoring counter-rotating terms, the Hamiltonian becomes effectively time independent:
\begin{equation}\label{Heff}
\hat H(\hat a,\hat a^\dagger)\simeq\Delta\hat a^\dagger\hat a+\lambda(\hat a^{\dagger2}+\hat a^2),
\end{equation}
where $\Delta=\omega_0-\omega_p$.  Integrating out the bath degrees of freedom, leads to a dissipative dynamics, described by the Lindblad equation for the reduced density matrix of the oscillator:
\begin{equation}\label{LinbladEq}
\partial_t\rho=\doublehat\fL\rho=-i[\hat H,\rho]+\gamma\Big(\hat a^2\rho\hat a^{\dagger2}-\frac{1}{2}\{\hat a^{\dagger2}\hat a^2,\rho\}\Big),
\end{equation}
where the Lindbladian `superoperator' $\doublehat\fL$ acts in the Hilbert space of operators and the dissipation strength is  $\gamma=\sum_s(\pi g_s^2/2\omega_s)\delta(\omega_0-\omega_s)$.

Both $\hat H$ and the jump operator $\hat a^2$ commute with $\hat P$, making Eq.~(\ref{LinbladEq}) invariant under $\hat a\to-\hat a$ on both the left and the right of $\rho$ independently.  Defining the left/right parity superoperators $\doublehat\fP_\pm$ to act by left/right multiplication by $\hat P$: $\doublehat\fP_+\rho=\hat P\rho$, $\doublehat\fP_-\rho=\rho\hat P$, this is made precise by the fact that both commute with the Lindbladian $[\doublehat\fP_\pm,\doublehat\fL]=0$.  In this case the system is said to posses the strong parity symmetry.  The simultaneous diagonalizability of $\doublehat\fL$ and $\doublehat\fP_-, \doublehat\fP_+$ means that there are at least two stationary density matrices.  If this symmetry is broken, there are {\em four} stationary density matrices, isomorphic to the Bloch sphere of of steady states \cite{Albert}.

Models of a parametrically driven oscillator linearly coupled to a bath have been studied extensively in, for example, \cite{Dykman1,Dykman2,Dykman3}.  Such models posses only weak symmetry and so can exhibit only classical, not quantum, bi-stability.  Models with master equations similar to (\ref{LinbladEq}) have been studied using methods based on operator formalism \cite{AlbertCats,Clerk} and quasi-probability distributions \cite{Minganti1,Minganti2}.  Experimentally, this is realizable in a pumped system of microwave cavities coupled via Josephson junction \cite{Devoret1, Devoret2}.

When the drive amplitude $\lambda$ is sufficiently large, the oscillator becomes unstable at the level of the Hamiltonian (\ref{Heff}).  To see this, it is convenient to go to Hermitian operators $\hat x,\hat p$, such that $\hat a=(\hat p-i(\Delta+2\lambda)\hat x)/\sqrt{(\Delta+2\lambda)}$, in terms of which Eq.~(\ref{Heff}) becomes $\hat H= \hat p^2+(\Delta^2-4\lambda^2)\hat x^2$.  This is a conventional harmonic oscillator so long as $\Delta>2\lambda$, but for $\Delta<2\lambda$ the potential flips upside down causing instability.  In the latter case the model is stabilized by the nonlinear dissipation.  This can be seen by, eg., taking  $\Delta=0$ and noticing that the Hamiltonian can be absorbed into the dissipative part of the Lindbladian by a linear shift of the jump operator:
\begin{equation}\label{LindbladEqDelta0}
\doublehat\fL\rho=\gamma\Big((\hat a^2-\phi_0^2)\rho (\hat a^{\dagger2}-\phi_0^{*2})-\frac{1}{2}\big\{(\hat a^{\dagger2}-\phi_0^{*2})(\hat a^2-\phi_0^2),\rho\big\}\Big),
\end{equation}
where $\phi_0=\sqrt{2\lambda/i\gamma}$.  
The two coherent states $\ket{\pm\phi_0}$ are anihilated by the shifted jump operators, so the {\em four} density matrices formed by their products $\ket{\pm\phi_0}\bra{\pm\phi_0}$ and $\ket{\pm\phi_0}\bra{\mp\phi_0}$ are all stationary.  These operators are exchanged by the right or left action of $\hat P$, meaning the parity symmetry is spontaneously broken.  The even and odd cat states $\ket{\pm}\propto\ket{\phi_0}\pm\ket{-\phi_0}$ form an orthonormal basis of the steady-state space.  Pure states made from their linear combinations form a Bloch sphere.

The model is zero-dimensional, so symmetry breaking cannot occur in the bulk of the parameter space.  When $\Delta$ is small but finite, the symmetry must be restored.  An exponentially small dissipative gap will open in the Lindbladian spectrum as the eigenvalues of the operators $\ket{\pm}\bra{\mp}$ move into the complex plain as a conjugate pair, denoted as $\Lambda$ and $\Lambda^*$.  The stationary-state space is reduced in dimension from four to two.  This phenomenon is non-perturbative in $\gamma$ and occurs due to instanton contributions to the Keldysh path integral.

To show this, it is useful to study the character-valued ``partition functions'' $\Tr(\doublehat\fP_\pm\exp(T\doublehat\fL))$ and $\Tr(\doublehat\fP_+\doublehat\fP_-\exp(T\doublehat\fL))$, where the capital Tr indicates that the trace is over a superoperator and $T\to\infty$ is the evolution time.  The action on the eigenvectors  $\fP_+\ket{-}\bra{\pm}=-\ket{-}\bra{\pm}$ means the their eigenvalues come with an additional sign in the $\fP_+$-twisted trace, and similarly for the other parity superoperators.  Assuming all nonzero eigenvalues of $\doublehat\fL$ have a finite real part, only the smallest eigenvalues contribute in the limit $T\to\infty$:
\begin{subequations}\label{CharValuedZ}
\begin{equation}
\Tr\big(\exp(T\doublehat\fL)\big)\simeq2+e^{\Lambda T}+e^{\Lambda^* T}\simeq4+T(\Lambda+\Lambda^*),
\end{equation}
\begin{equation}
                        \label{ImLambda}
\Tr\big(\doublehat\fP_\pm\exp(T\doublehat\fL)\big)\simeq \pm(e^{\Lambda T}-e^{\Lambda^* T})\simeq\pm T(\Lambda-\Lambda^*),
\end{equation}
\begin{equation}
                        \label{ReLambda} 
\Tr\big(\doublehat\fP_+\doublehat\fP_-\exp(T\doublehat\fL)\big)\simeq2-e^{\Lambda T}-e^{\Lambda^* T}\simeq-T(\Lambda+\Lambda^*),
\end{equation}
\end{subequations}
where the last approximate equalities indicate one instanton contributions.

Using the machinery of Ref.~\cite{Diehl}, the character-valued partition functions may be represented as coherent state Keldysh path integrals $\int\fdif\phi^\pm\fdif\bar\phi^\pm e^{iS}$ over the spaces of twisted loops in the Keldysh phase space,~\cite{Suplemental}. Here the fields $(\phi^\pm,\bar\phi^\pm)$ reside on the forward/backward part of the closed time contour and the action is given by
\begin{equation}
                        \label{action}
S=\int\limits_{-T/2}^{T/2}\!\dif t\, \big(\bar\phi^+ i\partial_t\phi^+-\bar\phi^- i\partial_t\phi^--H^++H^--iD\big),
\end{equation}
where $H^\pm=H(\phi^\pm,\bar\phi^\pm)$ from Eq.~(\ref{Heff}) and
\begin{equation}
D(\phi^\pm,\bar\phi^\pm)=\gamma\Big((\bar\phi^-\phi^+)^2-\frac{1}{2}(\bar\phi^+\phi^+)^2-\frac{1}{2}(\bar\phi^-\phi^-)^2\Big).
\end{equation}
Notice that the dissipator, $D$, breaks the time-reversal symmetry between the forward and backward parts of the contour.   
The strong Lindbladian parity symmetry is realized as a classical $\Z_2\times\Z_2$ symmetry of the Keldysh action: transforming either of the sets of fields independently $\phi^\pm\to-\phi^\pm$ leaves the action invariant.

We will consider a saddle point approximation to the character-valued partition functions described above.  For this, one must find solutions to the classical EoM with the appropriate boundary conditions for $T\to\infty$.  Expressing  the action (\ref{action}) in terms of the Keldysh rotated fields \cite{QFT&NonEq} $\phi^\pm=(\phi^\cl\pm \phi^\q)/\sqrt{2}$, one finds for the action: 
$S=\int\!\dif t\, \Big(\bar\phi^\q i\partial_t\phi^\cl+\bar\phi^\cl i\partial_t\phi^\q-K\Big)$,
where 
\begin{multline}
                        \label{KeldK}
K(\phi^\cl,\phi^\q,\bar\phi^\cl,\bar\phi^\q)=\Delta(\bar\phi^\q\phi^\cl+\bar\phi^\cl\phi^\q)+2\lambda(\bar\phi^\q\bar\phi^\cl+\phi^\q\phi^\cl)\\-i\frac{\gamma}{2}\big((\bar\phi^\cl\phi^\cl-\bar\phi^\q\phi^\q)(\bar\phi^\q\phi^\cl-\bar\phi^\cl\phi^\q)+4\bar\phi^\q\phi^\q\bar\phi^\cl\phi^\cl\big).
\end{multline}
The classical $\Z_2\times\Z_2$ symmetry realized on these fields is generated by $\phi^{\cl,\q}\to-\phi^{\cl,\q}$ and $\phi^\cl\leftrightarrow\phi^\q$.

The classical EoM are found by varying the action in each of the four fields. They acquire the Hamiltonian structure with the conserved ``energy'' $K$:
\begin{equation}\label{EoM}
i\partial_t\phi^\cl=\partial_{\bar\phi^\q}K,\ \ \ -i\partial_t\bar\phi^\cl=\partial_{\phi^\q}K,
\end{equation}
and the other EoM are given by exchanging labels $\cl\leftrightarrow\q$.  In addition to the trivial fixed point at the origin of the phase space, EoM admit additional nontrivial fixed points~\cite{Suplemental}.  When $\Delta=0$, these occur when either the $\cl$ or $\q$ fields are equal to $\pm\sqrt{2}\phi_0$ from Eq.~(\ref{LindbladEqDelta0}) while the complimentary fields are zero.  At these four fixed points the classical $\Z_2\times\Z_2$ symmetry is spontaneously broken.  Away from the $\Delta=0$ limit, the symmetry is restored quantum mechanically due to instanton solutions to the EoM, connecting the fixed points in the complexified phase space of the field configurations.

Indeed, EoM (\ref{EoM}) do {\em not} respect $\bar\phi^{\cl,\q}$ being complex conjugated of $\phi^{\cl,\q}$, but rather operate in an enlarged complexified phase space, where    
$\phi^{\cl,\q}$ and $\bar\phi^{\cl,\q}$ are regarded as {\em independent}  complex variables. This should be understood as  a complex deformation of the original integration contours over $\phi^{\cl,\q}, \bar\phi^{\cl,\q}$  to pass through  proper saddle-point field configurations. The classical motion, resulting from this complexification, takes place in the eight real-dimensional phase space spanned by the four holomorphic coordinates $(\phi^\cl,\phi^\q,\bar\phi^\cl,\bar\phi^\q)$ and their complex conjugates $(\phi^{\cl*},\phi^{\q*},\bar\phi^{\cl*},\bar\phi^{\q*})$. In this eight-dimensional space, $K+K^*$ acts as the Hamiltonian generating the motion and $K-K^*$ is an integral of motion.  In addition to this, there is a hidden symmetry generated by the holomorphic integral of motion $M$.  This can be found by looking for holomorphic polynomials that Poisson--commute with $K$:
\begin{equation}
K(\cev\partial_{\phi^\cl}\vec\partial_{\bar\phi^\q}-\cev\partial_{\bar\phi^\q}\vec\partial_{\phi^\cl}+\cev\partial_{\phi^\q}\vec\partial_{\bar\phi^\cl}-\cev\partial_{\bar\phi^\cl}\vec\partial_{\phi^\q})M=0.
\end{equation}
For $\Delta=0$ (see \cite{Suplemental} for a more general case), the lowest order solution (not equal to a multiple of $K$)  is given by:
\begin{eqnarray}
&&\!\! M\!\!=\!\!\Big[2\phi_0^2(\phi^{\cl2}\!+\!\bar\phi^{\cl2})\!+\!(\bar\phi^\cl\phi^\q\!-\!\bar\phi^\q\phi^\cl\!+\!2\bar\phi^\cl\phi^\cl)(\bar\phi^\q\phi^\cl\!+\!\bar\phi^\cl\phi^\q)\Big]\nonumber \\
&&\!\times\Big[2\phi_0^2(\phi^{\q2}\!+\!\bar\phi^{\q2})\!-\!(\bar\phi^\cl\phi^\q\!-\!\bar\phi^\q\phi^\cl\!-\!2\bar\phi^\q\phi^\q)(\bar\phi^\q\phi^\cl\!+\!\bar\phi^\cl\phi^\q)\Big].\nonumber 
\end{eqnarray}
As a consequence $K$, $M$, and their conjugates are four independent conserved quantities. This makes the corresponding eight-dimensional classical motion completely integrable.  

\begin{figure}
\begin{center}
\vskip .5cm
\scalebox{.45}{\includegraphics{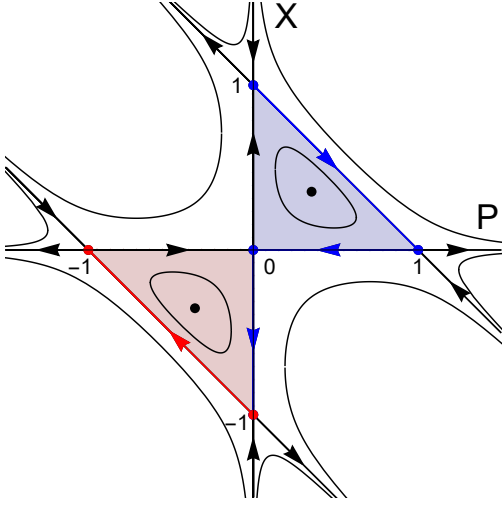}}
\caption{Phase portrait of the Hamiltonian $K(X,P)$, Eq.~(\ref{real-K}). Bold directed lines show separatrix trajectories $K=0$, thinner lines show examples of other trajectories, and dots show fixed points.  Instanton paths that contribute to the real and imaginary parts of $\Lambda$ highlighted in blue and red, respectively.    The first satisfies the boundary condition $X(-\infty)=-X(\infty)$ and $P(-\infty)=-P(\infty)$ and the second, $X(-\infty)=P(\infty)$ and $P(-\infty)=X(\infty)$. The shaded areas show the instanton action for each path.}
\label{PhasePortraitInst}
\end{center}
\end{figure}

In the limit $T\to\infty$, the instantons are given by the separatrix trajectories connecting the fixed points.  Since at the fixed points $K=M=0$, this is also the case for the separatrix trajectories and thus all four conserved quantities are equal to zero.  The condition $K+K^*=0$ dictates $\bar\phi^\cl=\phi^{\cl*}$ and $\bar\phi^\q=-\phi^{\q*}$, while the condition $M=0$ fixes $\phi^\cl=i\bar\phi^\cl$ and $\phi^\q=-i\bar\phi^\q$.  Each of these conditions defines a four-dimensional sub-space of the eight-dimensional phase space.  Their intersection is thus a 2D plane parametrized by the real coordinates $(X,P)$, with
\begin{equation}\label{Rescaling}
\phi^\cl=i\phi^{\cl*}=\sqrt\frac{4\lambda}{i\gamma}\,\,X; \qquad \bar\phi^\q=i\bar\phi^{\q*}=-\sqrt\frac{4i\lambda}{\gamma}\,\,P.
\end{equation}
At this invariant plain, the Keldysh action acquires an explicit Hamiltonian structure: 
\begin{equation}
                    \label{real-action}
iS=\frac{8\lambda}{\gamma}\int\limits_{-T/2}^{T/2}\dif t\Big(P\partial_tX-
K(X,P)\Big), 
\end{equation}
where the effective Hamiltonian is found by the substitution of Eq.~(\ref{Rescaling}) into Eq.~(\ref{KeldK}): 
\begin{equation}
                \label{real-K}
K(X,P)=2\lambda \, XP(1+P+X)(1-P-X).
\end{equation}
The parity symmetry manifests itself in the invariance vis-a-vis $X\leftrightarrow P$ exchange.   
Notice that Eq.~(\ref{real-action}) is a {\em real} time action.  Unlike in Hermitian quantum mechanics, there is no need to go to imaginary time to find connecting trajectories.  
The relevant fixed points in the new coordinates are $(\pm1,0)$ and $(0,\pm1)$.  As can be seen from the phase portrait for the Hamiltonian (\ref{real-K}), Fig.~\ref{PhasePortraitInst}, there are several instanton trajectories connecting the various fixed points.

\begin{figure}
\begin{center}
\scalebox{.5}{\includegraphics{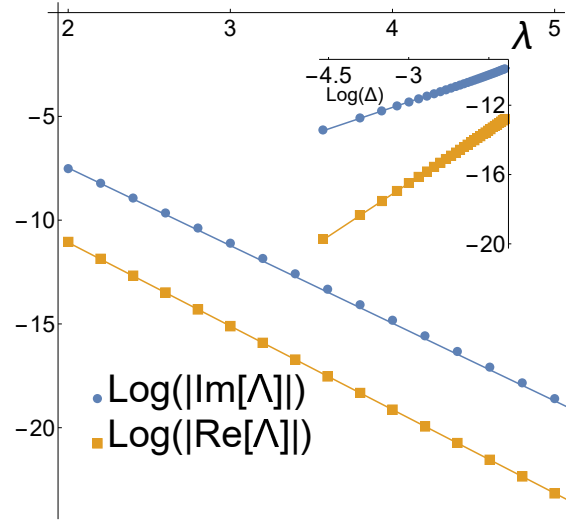}}
\caption{Imaginary and real parts of $\Lambda$ vs. the drive strength $\lambda$ with $\gamma=1$ and $\Delta=0.1$.  Best linear fit lines are plotted over the data and have slopes of $-3.7$ and $-4.0$, correspondingly.  The insert is log-log plot of $\Lambda$ vs. $\Delta$ for $\lambda=3$.  Best fits of the imaginary and real parts of $\Lambda$ come with the slopes of $1.04(\pm0.02)$ and $2.04(\pm 0.02)$.}
\label{Numerics}
\end{center}
\end{figure}

According to Eqs.~(\ref{ImLambda}), (\ref{ReLambda}) the imaginary and real parts of the smallest non-zero eigenvalue $\Lambda$ are determined by the one instanton contributions to the corresponding partition functions (the factor $T$ there originates from the freedom to choose the instanton position along the time axis). They are distinguished by the choice of  boundary conditions~\cite{Suplemental}, illustrated  in Fig.~\ref{PhasePortraitInst}.  Notably the action along all such paths is same:
\begin{equation}\label{exponent}
iS_\mathrm{inst}=\frac{8\lambda}{\gamma}\int\dif XP= -\frac{4\lambda}{\gamma}.
\end{equation}
This leads to  $\Lambda\sim \exp(iS_\mathrm{inst})=\exp(-4\lambda/\gamma)$.    
Numerical calculations (see Ref.~\cite{Suplemental} for details) of  $\Lambda$, shown in Fig.~\ref{Numerics},  are in a good agreement with this result.  The pre-exponential factors in the imaginary and real  parts appear to have different dependence on $\Delta$, with our numerics being consistent with: 
\begin{equation}
\mathrm{Im}\, \Lambda \sim \Delta\, e^{-4\lambda/\gamma}; \qquad 
\mathrm{Re}\, \Lambda \sim \Delta^2\, e^{-4\lambda/\gamma}. 
\end{equation}
The former represents the qubit precession frequency, while the latter -- its decoherence rate.    

To conclude: we have developed the instanton calculus to evaluate non-perturbative phenomena within the Lindbladian dynamics of the parametrically driven oscillator with the non-linear dissipation. It exhibits two oscillatory states, distinguished by their phases relative to the external drive, forming a qubit basis. Such qubit is, however, subject to a precession and decoherence due to the non-perturbative   
symmetry restoration. We found that the semiclassical description of this phenomenon relies on the real-time instanton trajectories in the complexified (and Keldysh doubled) phase space of the oscillator. This put the corresponding decoherence mechanism in the universality class of the {\em quantum activation} \cite{Dykman1,Dykman2,Dykman3}.

We are indebted to  Mark Dykman for numerous discussions. We also acknowledge helpful conversations with Victor Albert.  
This work was supported by the  NSF grant DMR-2037654.

\pagebreak
\widetext
\begin{center}
\textbf{\large{Supplemental Material for ``Qubit Dephasing and Symmetry Restoration through Real-Time Instantons"}}
\end{center}

\section{Equations of Motion Details}
The equations of motion for of the Keldysh action written in terms of the Keldysh-rotated fields are Hamilton's equation in the Keldysh Hamiltonian $K$:
\begin{subequations}\label{FullEoM}
\begin{equation}\label{EoMc}
i\partial_t\phi^\cl=\partial_{\bar\phi^\q}K=\Delta\phi^\cl+2\lambda\bar\phi^\cl-i\frac{\gamma}{2}\big(\bar\phi^\cl(\phi^{\cl2}+\phi^{\q2})-2\bar\phi^\q\phi^\q\phi^\cl+4\bar\phi^\cl\phi^\cl\phi^\q\big),
\end{equation}
\begin{equation}\label{EoMC}
-i\partial_t\bar\phi^\cl=\partial_{\phi^\q}K=\Delta\bar\phi^\cl+2\lambda\phi^\cl+i\frac{\gamma}{2}\big(\phi^\cl(\bar\phi^{\cl2}+\bar\phi^{\q2})-2\phi^\q\bar\phi^\q\bar\phi^\cl-4\bar\phi^\cl\phi^\cl\bar\phi^\q\big),
\end{equation}
\begin{equation}
i\partial_t\phi^\q=\partial_{\bar\phi^\cl}K=\Delta\phi^\q+2\lambda\bar\phi^\q-i\frac{\gamma}{2}\big(\bar\phi^\q(\phi^{\q2}+\phi^{\cl2})-2\bar\phi^\cl\phi^\cl\phi^\q+4\bar\phi^\q\phi^\q\phi^\cl\big),
\end{equation}
\begin{equation}
-i\partial_t\bar\phi^\q=\partial_{\phi^\cl}K=\Delta\bar\phi^\q+2\lambda\phi^\q+i\frac{\gamma}{2}\big(\phi^\q(\bar\phi^{\q2}+\bar\phi^{\cl2})-2\phi^\cl\bar\phi^\cl\bar\phi^\q-4\bar\phi^\q\phi^\q\bar\phi^\cl\big).
\end{equation}
\end{subequations}
Fixed points of the motion are determined by setting the left-hand side of each expression to zero.  There is a trivial fixed point in which all four fields are zero.  Additional fixed points can be determined by setting either the $\cl$ or the $\q$ fields equal to zero and solving the resulting algebraic equations.  For $\phi^\q=0=\bar\phi^\q$, this is:
\begin{equation}
\Delta\phi^\cl+2\lambda\bar\phi^\cl-i\frac{\gamma}{2}\bar\phi^\cl\phi^{\cl2}=0=\Delta\bar\phi^\cl+2\lambda\phi^\cl-i\frac{\gamma}{2}\bar\phi^{\cl2}\phi^\cl.
\end{equation}
One may look for solutions when $\phi^\cl=\phi_*=\bar\phi^{\cl*}$; there are two solutions:
\begin{equation}
\phi_*=\pm\sqrt{\frac{4\lambda}{\gamma}\sqrt{1-\Big(\frac{\Delta}{2\lambda}\Big)^2}}e^{i\varphi}, \ \ \ \cos(2\varphi)=-\frac{\Delta}{2\lambda},\ \ \ \sin(2\varphi)<0.
\end{equation}
Alternatively, one may look for solutions $-\phi^{\cl*}=\chi_*=\bar\phi^\cl$; there are also two solutions:
\begin{equation}
\chi_*=\pm\sqrt{\frac{4\lambda}{\gamma}\sqrt{1-\Big(\frac{\Delta}{2\lambda}\Big)^2}}e^{i\varphi}, \ \ \ \cos(2\varphi)=\frac{\Delta}{2\lambda},\ \ \ \sin(2\varphi)>0.
\end{equation}
Analogously, there are also fixed points for $\phi^\cl=0=\bar\phi^\cl$ and either $\phi^\q=\phi_*=\bar\phi^{\q*}$ or $-\phi^{\q*}=\chi_*=\bar\phi^\q$.  In the $\Delta\to0$ limit, $\phi_*=\pm\sqrt{2}\phi_0$ and $\chi_*=\pm\sqrt{2}\phi_0^*$ from (\ref{LindbladEqDelta0}).  As a consequence, the locations of the fixed points of the forward/backward fields $\phi^\pm$ match the steady-state coherent states.  The fixed points important for this problem are those for which $\phi^\cl=0=\bar\phi^\cl$,  $-\phi^{\q*}=\chi_*=\bar\phi^\q$ and $\phi^\q=0=\bar\phi^\q$, $\phi^\cl=\phi_*=\bar\phi^{\cl*}$.  When $\Delta=0$ all four of these fixed points exist in the two-dimensional plain in the complex Keldysh phase space on which $K+K^*=0=M$.

Note that one of the sets of fixed points occurs when $\bar\phi^\q\neq\phi^{\q*}$, where $(\cdot)^*$ is complex conjugate.  Similarly, any non-static solutions to the EoM require some complexification of the fields.  This is apparent when comparing Eq.s~(\ref{EoMc}) and (\ref{EoMC}): these equations are not complex conjugates of one another when $\bar\phi^{\cl,\q}=\phi^{\cl,\q*}$.  As a consequence, any non-static solutions will require $\phi^{\cl,\q}$ and $\bar\phi^{\cl,\q}$ to be regarded as independent complex variables.  The four equations of motion from Eq.s~(\ref{FullEoM}) and their four complex conjugates specify eight equations of motion on an eight real-dimensional complexified phase space with the four holomorphic and four anti-holomorphic coordinates $(\phi^\cl,\phi^\q,\bar\phi^\cl,\bar\phi^\q,\phi^{\cl*},\phi^{\q*},\bar\phi^{\cl*},\bar\phi^{\q*})$.  These EoM are Hamilton's equations generated by the Hamiltonian $K+K^*$.  The corresponding Poisson structure has a bracket with holomorphic and antiholomorphic parts:
\begin{eqnarray}
\{A,B\}=-iA\big(\cev\partial_{\phi^\cl}\vec\partial_{\bar\phi^\q}-\cev\partial_{\bar\phi^\q}\vec\partial_{\phi^\cl}+\cev\partial_{\phi^\q}\vec\partial_{\bar\phi^\cl}-\cev\partial_{\bar\phi^\cl}\vec\partial_{\phi^\q}-\cev\partial_{\phi^{\cl*}}\vec\partial_{\bar\phi^{\q*}}+\cev\partial_{\bar\phi^{\q*}}\vec\partial_{\phi^{\cl*}}-\cev\partial_{\phi^{\q*}}\vec\partial_{\bar\phi^{\cl*}}+\cev\partial_{\bar\phi^{\cl*}}\vec\partial_{\phi^{\q*}}\big)B=0.
\end{eqnarray}
The holomorphic part of the Poisson bracket kills any anti-holomorphic function and vice-versa.  As a consequence, holomorphic and anti-holomorphic functions always commute.  Moreover, two holomorphic functions $A$ and $B$ commute when
\begin{eqnarray}
0=A\big(\cev\partial_{\phi^\cl}\vec\partial_{\bar\phi^\q}-\cev\partial_{\bar\phi^\q}\vec\partial_{\phi^\cl}+\cev\partial_{\phi^\q}\vec\partial_{\bar\phi^\cl}-\cev\partial_{\bar\phi^\cl}\vec\partial_{\phi^\q}\big)B.
\end{eqnarray}
In such cases, there conjugates will also commute with one another, yielding a set of four commuting functions $A,B,A^*,B^*$.  It is this fact that explains why the complexified motion in the problem considered here is integrable.  There is an additional integral of motion $M$ commutes with $K$:
\begin{multline}
M=\bigg(2\phi_0^2(\phi^{\cl2}+\bar\phi^{\cl2})+(\bar\phi^\cl\phi^\q-\bar\phi^\q\phi^\cl+2\bar\phi^\cl\phi^\cl)\Big(\bar\phi^\q\phi^\cl+\bar\phi^\cl\phi^\q+\frac{2i\Delta}{\gamma}\Big)\bigg)\\
\bigg(2\phi_0^2(\phi^{\q2}+\bar\phi^{\q2})-(\bar\phi^\cl\phi^\q-\bar\phi^\q\phi^\cl-2\bar\phi^\q\phi^\q)\Big(\bar\phi^\q\phi^\cl+\bar\phi^\cl\phi^\q+\frac{2i\Delta}{\gamma}\Big)\bigg).
\end{multline}
This implies the existence of a total of three additional mutually Poisson-commuting functions $M,M^*,K-K^*$ alongside the Hamiltonian $K+K^*$.

\section{Keldysh Path Integral Details}
In the conventional construction of the Keldysh path integral, the object $1=\Tr(\exp(T\doublehat\fL)\rho)$ is converted into a functional integral by insertion of compete sets of coherent states on both the left and the right of the density matrix, as discussed in \cite{Diehl}.  Instead, we consider the trace of the time evolution superoperator $\Tr(\exp(T\doublehat\fL))$.  This trace can be expressed as the integral over the Hilbert-Schmidt operator norm of coherent states:
\begin{equation}\label{SuperTrace}
\Tr\big(\exp(T\doublehat\fL)\big)=\int\frac{\dif^2\phi_0^+\dif^2\phi_0^-}{\pi^2}e^{-\bar\phi_0^+\phi_0^+-\bar\phi_0^-\phi_0^-}\tr\Big(\ket{\phi_0^+}\bra{\phi_0^-}^\dagger\exp(T\doublehat\fL)\ket{\phi_0^+}\bra{\phi_0^-}\Big).
\end{equation}
The trace in right expression can be brought into the form of a path integral in a conventional way.  The remaining insertions $\ket{\phi_0^+}\bra{\phi_0^-}$ will set the boundary conditions at the beginning and end of the time contour.  At this step, the integral in front of the operator trace in (\ref{SuperTrace}) can be absorbed into the path integral measure to an integral over the loop space of the Keldysh phase space:
\begin{equation}
\Tr\big(\exp(T\doublehat\fL)\big)=\smashoperator{\int\limits_{\phi^\pm(-T/2)=\phi^\pm(T/2)}}\fdif\phi^\pm\fdif\bar\phi^\pm e^{iS}.
\end{equation}

The three character-valued partition functions $\Tr(\doublehat\fP_\pm\exp(T\doublehat\fL))$ and $\Tr(\doublehat\fP_+\doublehat\fP_-\exp(T\doublehat\fL))$ can be massaged into path integrals in the same way.  The parity superoperators will act to the left on the tensor of coherent states, replacing $\phi_0^\pm$ with $-\phi_0^\pm$ on the left or on the right, changing the boundary conditions on the fields.  The corresponding path integral must be taken over the space twisted loops on the Keldysh phase space.  The boundary conditions corresponding to the different twisting superoperators are:
\begin{subequations}
\begin{equation}
\fP_+\ \longleftrightarrow\ \phi^\pm(-T/2)=\mp\phi^\pm(T/2)
\end{equation}
\begin{equation}
\fP_-\ \longleftrightarrow\ \phi^\pm(-T/2)=\pm\phi^\pm(T/2)
\end{equation}
\begin{equation}
\fP_+\fP_-\ \longleftrightarrow\ \phi^\pm(-T/2)=-\phi^\pm(T/2)
\end{equation}
\end{subequations}
For the rescaled Keldysh-rotated fields $X$ and $P$ from (\ref{Rescaling}), these become:
\begin{subequations}\label{BC}
\begin{equation}
\fP_\pm\ \longleftrightarrow\ X(-T/2)=\mp P(T/2),\ P(-T/2)=\mp X(T/2)
\end{equation}
\begin{equation}
\fP_+\fP_-\ \longleftrightarrow\ X(-T/2)=-X(T/2),\ P(-T/2)=-P(T/2)
\end{equation}
\end{subequations}
Either of the first two can be used to isolate the imaginary part of the smallest finite Lindbladian eigenvalue $\Lambda$; the third is used to isolate the real part.

\section{Numerics}
Numerical calculations were obtained by truncating and then vectorizing the operator Hilbert space.  The Hilbert space was was truncated to finite dimension $N$; Boson creation/annihilation operators were replaced with $N\times N$ matrices.  Vectorization was done in the standard way by replacing simple tensor operators by vectors $\ket{\psi}\bra{\varphi}\to\ket{\psi}\otimes\ket{\varphi}$ and superoperators by operators acting on different tensor factors $A\cdot B\to A\otimes B^\mathrm{T}$.  After vectorization, the Lindbladian superoperator is transformed into an $N^2\times N^2$ matrix.  Its spectrum is obtained by exact diagonalization.  All numerical data shown was obtained using $N=50$.  Variation in the data was unnoticeable when going to larger values of $N$ for the range of parameters studied in the main text, meaning the results shown are effectively independent of $N$.  The low-lying part of the spectrum of the diagonalized truncated Lindbladian had two zero eigenvalues and two exponentially small eigenvalues; the plots in the main text show one of the two small eigenvalues.


\begin{thebibliography}{}
\bibitem{OpenQM}
H. Breuer and F. Petruccione, The Theory of Open Quantum Systems (Oxford University Press, 2002).

\bibitem{QuantumNoise}
C. W. Gardiner and P. Zoller, Quantum Noise (Springer-Verlag Berlin Heidelberg, 2004).

\bibitem{DrivenDissipativeLight}
C. Noh and D. G. Angelakis, Rep. Prog. Phys. {\bf 80}, 016401 (2017).

\bibitem{DrivenDissipativeColdAtoms}
S. Diehl, A. Micheli, A. Kantian, B. Kraus, H. P. Büchler, P. Zoller, Nature Phys. {\bf 4}, 878–883 (2008).

\bibitem{DiehlQuantumSim}
M. M{\"u}ller, S. Diehl, G. Pupillo, P. Zoller, Adv. At. Mol. Opt. Phys. {\bf 61}, 1 (2012).

\bibitem{Diehl} L. M. Sieberer, M. Buchhold, S. Diehl, Rep. Prog. Phys. {\bf 79}, 096001 (2016). 

\bibitem{NonAbelianSym}
Z. Zhang1, J. Tindall, J. Mur-Petit, D. Jaksch, B. Buča, J. Phys. A: Math. Theor. {\bf 53}, 215304 (2002).

\bibitem{AlbertGeom}
V. V. Albert, B. Bradlyn, M. Fraas, L. Jiang
Phys. Rev. X {\bf 6}, 041031 (2016).

\bibitem{NonEqSSKeldysh}
M. F. Maghrebi and A. V. Gorshkov, Phys. Rev. B {\bf 93}, 014307 (2016).

\bibitem{AlbertSym}
V. V. Albert and L. Jiang, Phys. Rev. A {\bf 89}, 022118 (2014).

\bibitem{LindbladSymBreak}
F. Minganti, A. Biella, N. Bartolo, C. Ciuti, Phys. Rev. A {\bf 98}, 042118 (2018).

\bibitem{Prosen}
B. Bu\v{c}a and T. Prosen, New J. Phys. {\bf 14}, 073007 (2012).

\bibitem{Clerk}
D. Roberts and A. A. Clerk, Phys. Rev. X {\bf 10}, 021022 (2020).

\bibitem{MultiDarkSpace}
R. A. Santos, F. Iemini, A. Kamenev, Y. Gefen, Nature Comm {\bf 11}, 5899 (2020).

\bibitem{QCompDrivenDissipative}
F. Verstraete, M. M. Wolf, J. I. Cirac, Nature Phys {\bf 5}, 633–636 (2009)

\bibitem{AlbertCats}
M. Mirrahimi, Z. Leghtas, V. V. Albert, S. Touzard, R. J. Schoelkopf, L. Jiang, M. H. Devoret, New J. Phys. {\bf 16}, 045014 (2014).

\bibitem{Albert} S. Lieu, R. Belyansky, J. T. Young, R. Lundgren, V. V. Albert A. V. Gorshkov, Phys. Rev. Lett. {\bf 125}, 240405 (2020).

\bibitem{DykmanClassical1}
M. I. Dykman, M. M. Millonas, V. N. Smelyanskiy. Phys. Lett. A, {\bf 195}, 53 (1994).

\bibitem{DykmanClassical2}
V. N. Smelyanskiy, M. I. Dykman, R. S. Maier. Phys. Rev. E, {\bf 55}, 2369 (1997).

\bibitem{Dykman1}
M. Marthaler and M. I. Dykman, Phys. Rev. A {\bf 73}, 042108 (2006).

\bibitem{Dykman2}
M. I. Dykman, ``Periodically modulated quantum nonlinear oscillators," in ``Fluctuating Nonlinear Oscillators. From nanomechanics to quantum superconducting circuits", ed. M. I. Dykman (Oxford University Press 2012), pp. 165--197.

\bibitem{Dykman3}
M. I. Dykman, M. Marthaler, and V. Peano, Phys. Rev. A 83, 052115 (2011).

\bibitem{Minganti1}
F. Minganti, N. Bartolo, J. Lolli, W. Casteels, C. Ciuti, Sci. Rep. {\bf 6}, 26987 (2016).

\bibitem{Minganti2}
N. Bartolo, F. Minganti, W. Casteels, C. Ciuti, Phys. Rev. A {\bf 94}, 033841 (2016).

\bibitem{Devoret1}
S. Touzard, A. Grimm, Z. Leghtas, S. O. Mundhada, P. Reinhold, C. Axline, M. Reagor, K. Chou, J. Blumoff, K. M. Sliwa, S. Shankar, L. Frunzio, R. J. Schoelkopf, M. Mirrahimi, M. H. Devoret, Phys. Rev. X {\bf 8}, 021005 (2018).

\bibitem{Devoret2}
Z. Leghtas, S. Touzard, I. M. Pop, A. Kou, B. Vlastakis, A. Petrenko, K. M. Sliwa, A. Narla, S. Shankar, M. J. Hatridge, M. Reagor, L. Frunzio, R. J. Schoelkopf, M. Mirrahimi, M. H. Devoret, Science {\bf 347}, 6224 pp. 853-857 (2015).

\bibitem{Suplemental}
See online Supplemental Material for this article.

\bibitem{QFT&NonEq}
A. Kamenev, Field Theory of Non-Equilibrium Systems (Cambridge University Press, 2011)


\end{thebibliography}
\end{document}